\documentclass[aps,prl,floats,superscriptaddress,floatfix,twocolumn,fleqn,longbibliography]{revtex4-2}

\usepackage[utf8]{inputenc}
\usepackage{dcolumn}
\usepackage{amsbsy}
\usepackage{amsmath}
\usepackage{orcidlink}
\usepackage{amssymb}
\usepackage{braket}
\usepackage{xcolor}
\usepackage{float}
\usepackage{graphicx}
\graphicspath{{./}{./Figs/}} 
\usepackage{bm}

\usepackage{hyperref}

\newcommand{\beq}{\begin{eqnarray}}
\newcommand{\eeq}{\end{eqnarray}}
\newcommand{\bea}{\begin{eqnarray}}
\newcommand{\eea}{\end{eqnarray}}

\def\<{\langle}
\def\>{\rangle}

\newcommand{\sectA}[1]
{
	\addtocounter{section}{1}
	\setcounter{subsection}{0}

	\pdfbookmark[1]{\thesection. \ #1}{sect.\thesection}
	{\Large\bf $=\!=\!=\!=\!=\!=\;$ [\thesection] \ #1}
	\nopagebreak
	\vspace*{3mm}
}
\renewcommand{\section}{\sectA}

\newcommand{\hide}[1]{}  %{{\textcolor{red}{[hide]}}}

\begin{document}

\title{Coexistence of Anderson Localization and Quantum Scarring in Two Dimensions}
\author{Fartash Chalangari \orcidlink{0009-0001-7182-204X}}
\affiliation{Computational Physics Laboratory, Tampere University, P.O. Box 600, FI-33014 Tampere, Finland}
\author{Anant Vijay Varma \orcidlink{0000-0002-7610-6317}}
\affiliation{Computational Physics Laboratory, Tampere University, P.O. Box 600, FI-33014 Tampere, Finland}
\author{Joonas Keski-Rahkonen \orcidlink{0000-0002-7906-4407}}
\affiliation{Computational Physics Laboratory, Tampere University, P.O. Box 600, FI-33014 Tampere, Finland}
\affiliation{Department of Physics, Harvard University, Cambridge, Massachusetts 02138, USA}
\affiliation{Department of Chemistry and Chemical Biology, Harvard University, Cambridge, Massachusetts 02138, USA}
\author{Esa Räsänen \orcidlink{0000-0001-8736-4496}}
\affiliation{Computational Physics Laboratory, Tampere University, P.O. Box 600, FI-33014 Tampere, Finland}

\date{\today}

\begin{abstract}
We investigate finite two-dimensional disordered systems with periodic confinement. At low energies, eigenstates exhibit strong Anderson localization, while at higher energies a subset of states exhibits variational scarring with anisotropic intensity patterns that deviate from random-wave expectations. Scaling theory predicts that in two dimensions all eigenstates localize in the large-system-size limit, yet the energy-dependent localization length and finite-size effects allow these regimes to coexist. We demonstrate that this coexistence produces distinct, robust signatures in both spatial intensity patterns and spectral statistics that are directly observable in mesoscopic electronic, photonic, and cold-atom systems.
\end{abstract}

\maketitle

%%%%%%%%%%%%%%%%%%%%%%%%%%%%%%%%%%%%%%%%%%%%%%%

\subsection{I.\quad Introduction} 
\noindent
In quantum systems, the breakdown of ergodicity can manifest in several distinct forms, from Anderson localization (AL) in non-interacting disordered settings~\cite{Anderson1958,Abrahams1979} to weakly non-ergodic structures that defy random‑wave expectations~\cite{DeTomasi2020Multifractality,Altshuler2016Nonergodic,Das2025Emergent,Xu2023NEE}, such as various species of quantum scars discussed further below. In the former, a quenched disorder localizes single-particle eigenstates in one and two dimensions, and suppresses transport~\cite{Billy2008,Roati2008} -- a principle that has evolved into an essential tenet of mesoscopic physics~\cite{Anderson_localization_book}. 

The latter, i.e., quantum scarring presents another archetypic example of non-ergodic behavior in single-particle systems in the form of atypical wavefunctions whose probability densities concentrate along classical periodic orbits (POs)~\cite{Heller_book,Heller_phys.rev.lett_53_1515_1984}. In this scenario, scarred states arise in non-integrable bounded systems where unstable POs embedded in a chaotic sea leave enhanced imprints on individual eigenstates, a mechanism rooted in constructive interference along quantized POs. Related manifestations sharing a similar semiclassical backbone include superscars associated with families of marginally stable POs in pseudointegrable systems~\cite{Bogomolny_2004}, as well as relativistic scars in Dirac materials where a Berry phase modifies the semiclassical quantization condition~\cite{Huang_phys.rev.lett_103_054101_2009}. These types of scars have been reported, e.g., in billiard-like platforms (see, Refs.~\cite{bogomolny_phys.rev.lett_97_254102_2006,Noeckel1997,Bogomolny2012} and Ref.~\cite{Ge_nature_635_841_2024}, respectively).

Variational scarring, on the other hand, operates through a fundamentally different mechanism in comparison to the conventional scarring~\cite{Heller_book,Heller_phys.rev.lett_53_1515_1984}, albeit a superficial resemblance in appearance. Specifically, weak (spatially local) perturbations couple states within degenerate manifolds of POs of the corresponding clean, structured periodic system, and degenerate perturbation theory selects anisotropic superpositions that extremize their overlap with the perturbation landscape~\cite{keski-rahkonen_phys.rev.lett_123_214101_2019,Selinummi2024,Chalangari2025,keskirahkonen_phys.rev.e_112_L012201_2025}. Crucially, whereas this type of disorder generally destroys conventional scars, it is an central ingredient for generating variational scars, directly visualized in semiconductor~\cite{Luukko_sci.rep_6_37656_2016, keski-rahkonen_phys.rev.lett_123_214101_2019} and graphene quantum wells~\cite{keskirahkonen_phys.rev.e_112_L012201_2025}.

Generally speaking, the localization length \(\xi(E)\) increases rapidly with energy \(E\)  two-dimensional (2D) disordered systems, such that finite samples of linear size \(L\) may realize both \(L \ll \xi(E)\) and \(L \gg \xi(E)\) regimes~\cite{RMTinQT}. 
In this crossover window, high-energy states can appear delocalized and display Wigner–Dyson level statistics~\cite{Mehta_book,PhysRevB103104201}, often regarded as a spectral signature of quantum chaos~\cite{Stockmann_book,Anderson_localization_book, Heller_book}. However, according to scaling theory~\cite{Abrahams1979}, a true metallic phase is excluded in the infinitely large size limit for 2D non-interacting systems with time-reversal symmetry, implying that all states localize for arbitrarily weak uncorrelated disorder. Short-range correlations in the disorder do not invalidate this conclusion for the most part, but rather renormalize the effective disorder strength and can modify \(\xi(E)\) by orders of magnitude~\cite{Conley2014,Monsarrat2022}. This energy dependence of the localization length naturally leads to finite-size crossover regimes in which localized and extended features may coexist.

In contrast, semiclassical propagation through an unbounded weakly disordered landscape can exhibit branched flow, where smooth random refraction generates caustics and filamentary intensity patterns~\cite{Daza2021,Heller2021,Heller_book}. This phenomenon has been observed, for instance, in 2D electron gases~\cite{Topinka2001} and optical media~\cite{laserbranch}. However, finite geometries qualitatively modify this picture: repeated returns of classical trajectories to the same regions of phase space enable coherent buildup along preferred paths. Therefore, finite clean systems whose bounded geometry supports families of classical POs support eigenstates that can exhibit enhanced probability density along these trajectories, interpreted as precursors of quantum scarring. Upon introducing local perturbations, the variational mechanism described above selects superpositions concentrated along specific POs, giving rise to variational scars~\cite{Luukko_sci.rep_6_37656_2016}.

The orientation of these scarred states tends to be selected variationally, either maximizing or minimizing the overlap with the effective impurity landscape~\cite{simo,keski-rahkonen_phys.rev.b_97_094204_2017,keski-rahkonen_phys.rev.lett_123_214101_2019,Luukko_phys.rev.lett_119_203001_2017}. This aspect follows from standard degenerate perturbation theory for a finite size set of nearly degenerate eigenstates of the unperturbed system, from which the perturbed system selects linear combinations extremizing the impurity potential. These features signal a form of weak ergodicity breaking that is distinct from typical localization phenomena, such as Anderson~\cite{Anderson1958,Abrahams1979, Anderson_localization_book}, dynamical~\cite{Casati_book, Izrailev_phys.rep_196_299_1990, Guarneri_Phys.Rev.Lett_113_174101_2014} or many-body localization~\cite{Abanin_rev.mod.phys_91_021001_2019}. 

Despite the progress on the frontiers above, the manner in which finite-size effects and continuum disorder conspire to yield coexisting non-ergodic regimes across the spectrum in realistic systems remains elusive. To address this issue, we here investigate a finite continuum Hamiltonian in two dimensions with prototypical disorder caused by a set of Gaussian potential bumps serving as a proxy for local impurities. Our analysis reveals regimes that separate across the energy spectrum: (i) conventional AL eigenstates dominating the low-energy sector, (ii) fully ergodic delocalized states following random-matrix-theory statistics in the sense of the Bohigas-Giannoni-Schmit conjecture~\cite{Bohigas_phys.rev.lett_52_1_1984}, while (iii) a set of high-energy scarred states exhibiting intensity patterns that deviate from random-wave predictions of the Berry conjecture~\cite{Berry_j.phys.a_10_2083_1977}. Representative eigenstates are shown in Fig.~\ref{fig:crossover}(c) and in the supplemental material (SM)~\cite{SupplementalMaterial} Fig.~\ref{fig:stateSamples}, illustrating behavior ranging from Anderson localization [Fig.~\ref{fig:crossover}(c-i)] to ergodic delocalization [Fig.~\ref{fig:crossover}(c-ii)] and various degrees of scarring [Figs.~\ref{fig:crossover}(c-iii),(c-iv)]. These regimes are distinguished employing complementary observables, including spatial profiles, inverse participation ratios (IPR), multifractal spectra, and deviations from universal level statistics.

%%%%%%%%%%%%%%%%%%%%%%%%%%%%%%%%%%%%%%%
\subsection{II.\quad Model Setting} 
We consider a 2D potential profile $V_{\mathrm{ext}}(\mathbf{r})$, confined by Dirichlet hard-wall boundaries to a region of size \(L\times L\), where \(L\) denotes the linear system size, i.e., the physical length of the square domain in each spatial direction, together with the effect of randomly scattered Gaussian impurities $V_{\text{imp}}(\mathbf{r})$. By utilizing the imaginary-time propagation method~\cite{LuukkoP.J.J.2013Itpc}, we compute the lowest $\sim 4\times 10^3-10^4$ eigenstates of the time-independent Schr{\"o}dinger equation \(H\psi_n(\mathbf{r})=E_n\psi_n(\mathbf{r})\) with
\begin{equation}
H = -\frac{1}{2}\nabla^2 + V_{\mathrm{ext}}(\mathbf{r}) + V_{\mathrm{imp}}(\mathbf{r}),
\label{eq:H}
\end{equation}
in atomic units (a.u.). In the following, eigenenergies $E_n$ are normalized as \(\tilde{E}_n = (E_n - E_{\min})/(E_{\max} - E_{\min})\), where $E_{\min}$ is the energy of the ground state and $E_{\max}$ is the highest eigenenergy obtained numerically.

\begin{figure}[!t]
    \centering
    \includegraphics[width=\columnwidth]{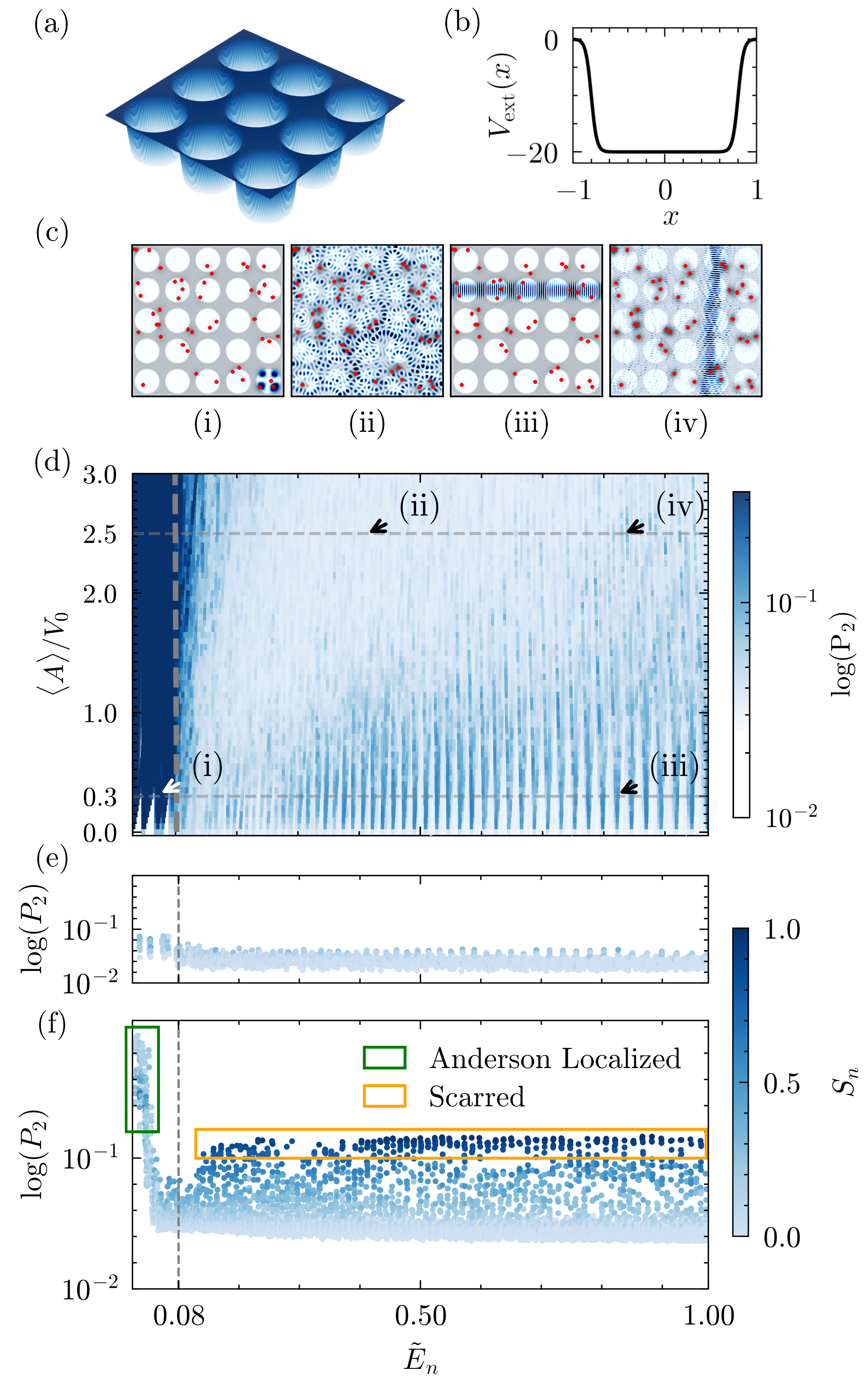}
    \caption{Energy--disorder diagram and coexistence of localization mechanisms. (a) Three-dimensional schematic of a $3\times3$ subset of the periodic external potential $V_{\mathrm{ext}}(\mathbf r)$ showing circular Fermi wells and surrounding barriers. (b) One-dimensional cross section of a single well.
    (c) Representative eigenstates illustrating (i) Anderson-localized, (ii) delocalized, (iii) strongly scarred, and (iv) weakly scarred states.
    (d) Disorder strength $\langle A\rangle/V_0$ vs normalized energy $\tilde{E}_n$, with color indicating $\log(\mathrm{P}_2)$; the gray dashed line marks the scaled well-depth ($\tilde{E}=0.08$). Simulations use $r_0=0.8$, $d=0.03$, $V_0=20$, $a=2$, and $L=10$ (a.u.). (e,f) Inverse participation ratio $P_2$ as a function of normalized energy for clean and disordered lattices, using the same lattice parameters as in panel (d). In panel (f), the disorder strength is fixed to $\langle A\rangle/V_0=0.3$, with the color indicating the scar metric $S_n$.}
    \label{fig:crossover}
\end{figure}

Our model Hamiltonian without \(V_{\text{imp}}\), has been considered to be non-integrable in Liouville sense \cite{Ott_2002}. Therefore, the corresponding classical dynamics is predominantly chaotic~\cite{Daza2021, klages_phys.rev.lett_122_064102_2019, graf_entropy_26_492_2024, Toivonen2025}, ensuring that no separable structure constrain the dynamics. The considered potential \(V_{\text{ext}}\) is a 2D array of wells, which are circular and described by the symmetrized Fermi profile
\begin{equation}
    V_{\text{ext}}(\mathbf{r}) = V_0 \frac{ \coth{\Big(\frac{r_0}{2d}\Big)}  \sinh(r_0/d)}{\cosh(r/d) + \cosh(r_0/d)},
\end{equation}
where \(r_0\), \(d\), and \(V_0\) set the well radius, edge smoothness, and well-depth~\cite{stefanPRB}, respectively. Throughout this work, we fix the Fermi potential parameters to \(r_0=0.8\), \(d=0.03\), \(V_0=20\), and spatial period \(a=2\) (all in a.u.), chosen to qualitatively match earlier experimental conditions~\cite{fermiexp1,fermiexp2}. A schematic of the external potential is shown in Fig.~\ref{fig:crossover}(a,b). Panel (a) presents a three-dimensional profile of a $3\times3$ portion of the periodic array, showcasing the circular potential wells and surrounding confinement barriers. Panel (b) further illustrates a one-dimensional cross section through a single well.

\begin{figure*}[!t]
    \centering
    \includegraphics[width=0.9\textwidth]{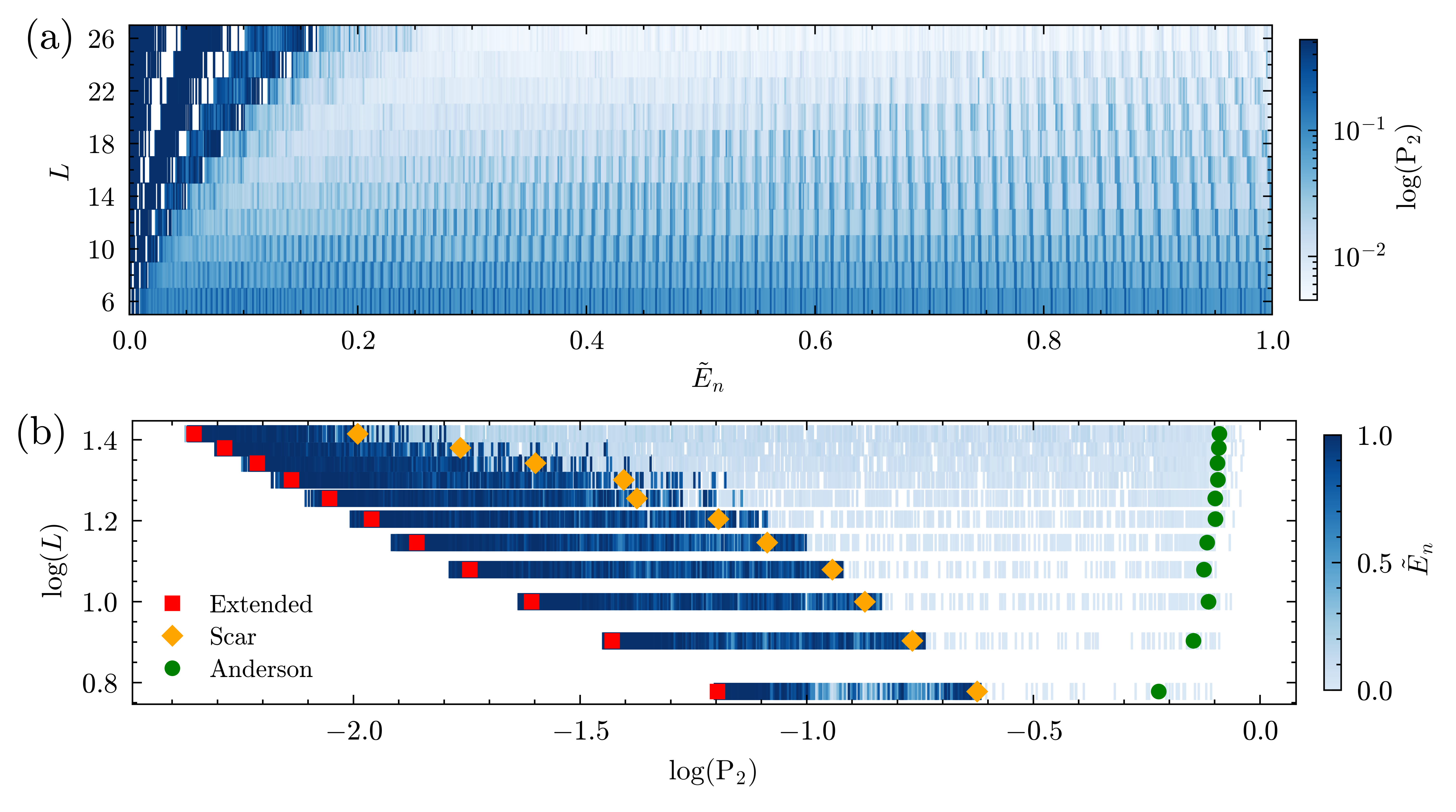}    
    \caption{Mesoscopic scaling analysis. (a) Normalized energy $\tilde{E}_n$ versus linear system size $L$ (in a.u.) for systems with fixed impurity density $\rho=0.4$ and disorder strength $\langle A\rangle/V_0=0.3$. The color scale shows $\log(\mathrm{P}_2)$. (b) Log-log plot of $\mathrm{P}_2$ versus $L$ for the same systems, with color indicating normalized energy $\tilde{E}_n$. Overlaid symbols show class-averaged $\langle \log(\mathrm{P}_2)\rangle$ for Anderson-localized (green), extended (red), and scarred (yellow) states, illustrating distinct finite-size scaling trends.}
\label{fig:scaling}
\end{figure*}

Disorder is introduced via randomly distributed Gaussian bumps as
\begin{equation}
V_{\text{imp}}(\mathbf{r})=\sum_i A_i \exp\!\left[-\frac{|\mathbf{r}-\mathbf{r}_i|^2}{2\sigma_i^2}\right],
\end{equation}
which is a standard model for imperfections in quantum wells~\cite{hirose_phys.Rev.B_63_075301_2001,hirose_phys.rev.B_65_193305_2002}. 
The impurity positions $\mathbf{r}_i$ are drawn from a uniform distribution over the system area, yielding an average impurity density $\rho$ per unit area. The amplitudes $A_i$ and widths $\sigma_i$ are sampled from uniform distributions centered around their respective mean values $\langle A \rangle$ and $\langle \sigma \rangle$, with a relative spread of $10\%$ about the mean. Throughout this work, the average impurity width is fixed at $\langle \sigma \rangle = 0.2$. The dimensionless ratio $\langle A \rangle / V_0$ characterizes the overall disorder strength.

%%%%%%%%%%%%%%%%%%%%%%%%%%%%%%%%%%%%%%%%%%%%
\subsection{III.\quad Coexistence of Non-ergodic Regimes} 

We begin by assessing the spectral and wavefunction properties of our system through the generalized IPR for a normalized eigenfunction \(\psi(\mathbf r)\) defined as
\begin{equation}
\mathrm{P}_q = \int d^2\mathbf r\, |\psi(\mathbf r)|^{2q} \sim L^{-D_q(q-1)},
\label{eq:scaling}
\end{equation}
which subsequently determines a generalized fractal dimension \(D_q\)~\cite{Girvin_Yang_2019}. We here restrict our attention to the typical second-order moment of \(q=2\) that is taken as a quantitative benchmark to classify the different spectral regimes.

Figure~\ref{fig:crossover}(d) shows an expected localization–delocalization crossover in the disorder–energy plane for a system with fixed size \(L\) and impurity density \(\rho=0.4\). The colormap shows logarithm of IPR, and the gray vertical dashed-line marks the scaled well-depth energy \(V_0\). At energies below and near the well-depth, \(E_n \lesssim V_0\), eigenstates are dominated by potential confinement and increasing disorder strength \(\langle A \rangle / V_0\) disrupts inter-well resonances and strongly localizes the wavefunctions (see a representative example in Fig.\ref{fig:crossover}(c-i)).
For these states, the angularly averaged density decays exponentially from the localization center, \(|\psi(r)|^2 \propto e^{-2r/\xi_{\mathrm{tail}}}\) (see Fig.~\ref{fig:exp_decay} in the SM~\cite{SupplementalMaterial}), aligning with the conventional Anderson localization scenery. 

In contrast, at energies \(E_n \gtrsim V_0\), the IPR landscape exhibits two quantitatively distinct regimes. The intermediate-energy limit consists of very low-IPR eigenstates in which scarring is essentially absent [see Fig.~\ref{fig:crossover}(c-ii)]. The localization length associated with these states exceeds the system size, and the de Broglie wavelength \(\lambda(E) \sim 2\pi/\sqrt{2(E-\langle V\rangle)}\), where \(\langle V\rangle\) denotes the expectation value of the total potential energy, remains too large to resolve the underlying periodic structure~\cite{Heller_book}. Consequently, the eigenstates behave as extended ergodic modes without pronounced directional features.

Upon further increasing the energy, the characteristic wavelength $\lambda(E)$ becomes sufficiently small to resolve the periodic confinement. Within this regime, degenerate subsets exist and can conform to variational scarred states when breaking the degeneracy. These scarred states appear as anisotropic outliers embedded within the delocalized background [see Fig.~\ref{fig:crossover}(c-iii)]. As the disorder strength increases, both the low- and intermediate-energy windows [Fig.~\ref{fig:crossover}(c-i) and (c-ii)] broaden, pushing the scar-bearing region to progressively higher energies. In the strong disorder limit, \(\langle A \rangle / V_0 \gtrsim 1\), the persisting scars are weakened [see Fig.~\ref{fig:crossover}(c-iv)] and occupy a diminishing fraction of the accessible spectrum, consistent with previous observations~\cite{KeskiRahkonen2019b,simo}.

The coexistence of the two distinct regimes described above is further clarified in Fig.~\ref{fig:crossover}(e,f), which show $\log(\mathrm{P}_2)$ as a function of normalized energy. Panel (e) corresponds to the clean periodic system and exhibits a narrow band of low-IPR states across the spectrum, underscoring largely ergodic eigenstates. When the disorder is introduced [panel (f)], the spectrum separates into two distinct high-IPR populations. Here we fix the disorder strength to \(\langle A\rangle/V_0=0.3\), chosen as a representative point within the scar-supporting region identified in Fig.~\ref{fig:crossover}(d). At low energies, large-IPR eigenstates correspond to Anderson-localized states, while a set of high-IPR outliers emerges within an otherwise delocalized background at higher energies. These states display pronounced anisotropy and are identified as scarred modes, as confirmed by the scar metric \(S_n\) encoded in the color scale (see SM~\cite{SupplementalMaterial} for details).

More specifically, the established metric quantifies the degree to which an eigenstate concentrates along row or column channels corresponding to POs in the finite periodic system. It combines probability weight and an elongation factor derived from probability-weighted spatial variances within stripe regions aligned with these channels, yielding stripe scores \(S^{\mathrm{row}}_j\) and \(S^{\mathrm{col}}_k\). The scar metric is subsequently defined as the maximum stripe score over all rows and columns,
\(S_n=\max(\max_j S^{\mathrm{row}}_j,\max_k S^{\mathrm{col}}_k)\). In particular,
large values of \(S_n\) indicate channel-like anisotropic states characteristic of variational scarring, whereas AL and ergodic states yield small values.

For the scaling analysis, we compute a fixed set of \(10^4\) eigenstates for each system size. Figure~\ref{fig:scaling}(a) shows how the spatial structure of the eigenstates, quantified by IPR, evolves with system size for fixed disorder strength \(\langle A \rangle / V_0 = 0.3\) and impurity density \(\rho = 0.4\). At energies \(E_n \lesssim V_0\), AL states occupy an increasingly large fraction of the sampled spectrum as \(L\) increases. This trend is follows the scaling theory of localization, which predicts that increasing \(L\) drives more states into the regime \(L/\xi(E) \gg 1\) at fixed energy.

At energies \(E_n \gtrsim V_0\), the spectrum contains two regimes, as already discussed in the context of Fig.~\ref{fig:crossover}. At intermediate-energies, the eigenstates are predominantly extended possessing very low IPR values, with no scarring, because the de Broglie wavelength remains too large to resolve the periodic potential structure. As the energy is increased further, the semiclassical mechanism outlined above comes into play: Degenerate manifolds associated with families of POs emerge, and local disorder can mix these states to produce variational scars. As a result, the high-energy region contains a mixture of ergodic and scarred eigenstates.

In 2D-systems, the mean level spacing scales as \(L^{-2}\). A fixed set of \(10^4\) resolved eigenstates hence spans a progressively narrower absolute energy window as the system size increases. In other words, larger systems require increasingly more eigenstates in order to reach the high-energy regime where the periodic structure becomes resolved and scars appear. This scaling can also be understood from Weyl's law~\cite{Girvin_Yang_2019,Gutzwiller_book}: the cumulative number of states below energy \(E\) follows \(N(E) \simeq (L^2/4\pi)E\), implying that the number of eigenstates required to reach a fixed characteristic energy grows proportionally to \(L^2\).

Figure~\ref{fig:scaling}(b) shows the same data as in panel (a), presented on a log--log scale of IPR versus the linear system size \(L\), with color encoding the normalized eigenenergy \(\tilde{E}_n\). To distinguish representative spectral classes in a reproducible fashion, we employ an energy-conditioned classification based on the IPR. In this manner, AL states are selected from the low-energy sector as eigenstates belonging to the high-IPR tail of the distribution. Extended states are chosen from the high-energy sector as the lower quintile of the IPR distribution, parallel to spatially ergodic scaling, while scarred states are identified as high-energy, high-IPR outliers evidencing additionally pronounced spatial anisotropy. Class-averaged values are evaluated separately for each category.

We find that a pronounced quantitative signature of coexistence is simultaneously present within the same spectral window and disorder realization, revealing three distinct finite-size scaling branches indicated by fractal dimensions \(D_2 \simeq 0\), \(D_2 \simeq 2\), and \(0 < D_2 < 2\). As stated in Eq.~\ref{eq:scaling} for 2D systems, ergodic states are expected to follow \(\mathrm{P}_2 \propto L^{-2}\), corresponding to an effective fractal dimension \(D_2 \simeq 2\), whereas AL states exhibit negligible \(L\)-dependence once \(L \gg \xi(E)\), yielding \(D_2 \simeq 0\). The extracted scaling exponent for the extended class is approximately \(-1.8\), close to the ideal value of \(-2\); the slight deviation reflects finite-size corrections and the restricted spectral window sampled at each \(L\), like discussed in Ref.~\cite{Girvin_Yang_2019} Conversely, AL states (green markers) display nearly vertical scaling consistent with \(\mathrm{P}_2 \sim \xi^{-2}\). Scarred states (yellow markers) fall into the class of intermediate and size-dependent behavior. For small systems, their mean scaling appears approximately linear on the log-log plot, with an effective exponent between \(0\) and \(2\), reflecting anisotropic support and partial ergodicity breaking.

As \(L\) increases, the scaling trend for scars gradually bends toward the ergodic branch. The drift arises because variational scars in this energy range rely on finite-size periodic structure; as the system grows, the density of competing states increases and the scar-carrying manifolds shift to higher energies outside the numerically sampled window. This fact does not preclude the existence of strong high-energy scars (e.g., superwire-like modes~\cite{Daza2021,graf_entropy_26_492_2024}) in larger periodic systems, but rather implies a mesoscopic crossover governed by the interplay between localization length and system size.

Heuristically speaking, the coexistence observed above arises from the interplay between the energy-dependent localization length \(\xi(E)\), the finite system size \(L\), and the underlying periodic confinement, which supports families of classical POs and associated degenerate manifolds. At low energies, the condition \(L \gg \xi(E)\) leads to AL states, whereas at higher energies \(L \lesssim \xi(E)\) allows states to extend across the sample. When the de Broglie wavelength becomes comparable to the lattice spacing, the periodic confinement becomes semi-classically resolved and families of POs emerge. This advent then gives rise to degenerate manifolds of high-energy eigenstates, whose structure and semiclassical quantization are well described within established frameworks~\cite{Tanner_1997}. In this regime, local disorder mixes states within these manifolds, and the resulting coherent superpositions concentrate probability density along specific orbit directions, producing anisotropic variational scars.

\begin{figure}[!t]
    \centering
    \includegraphics[width=\columnwidth]{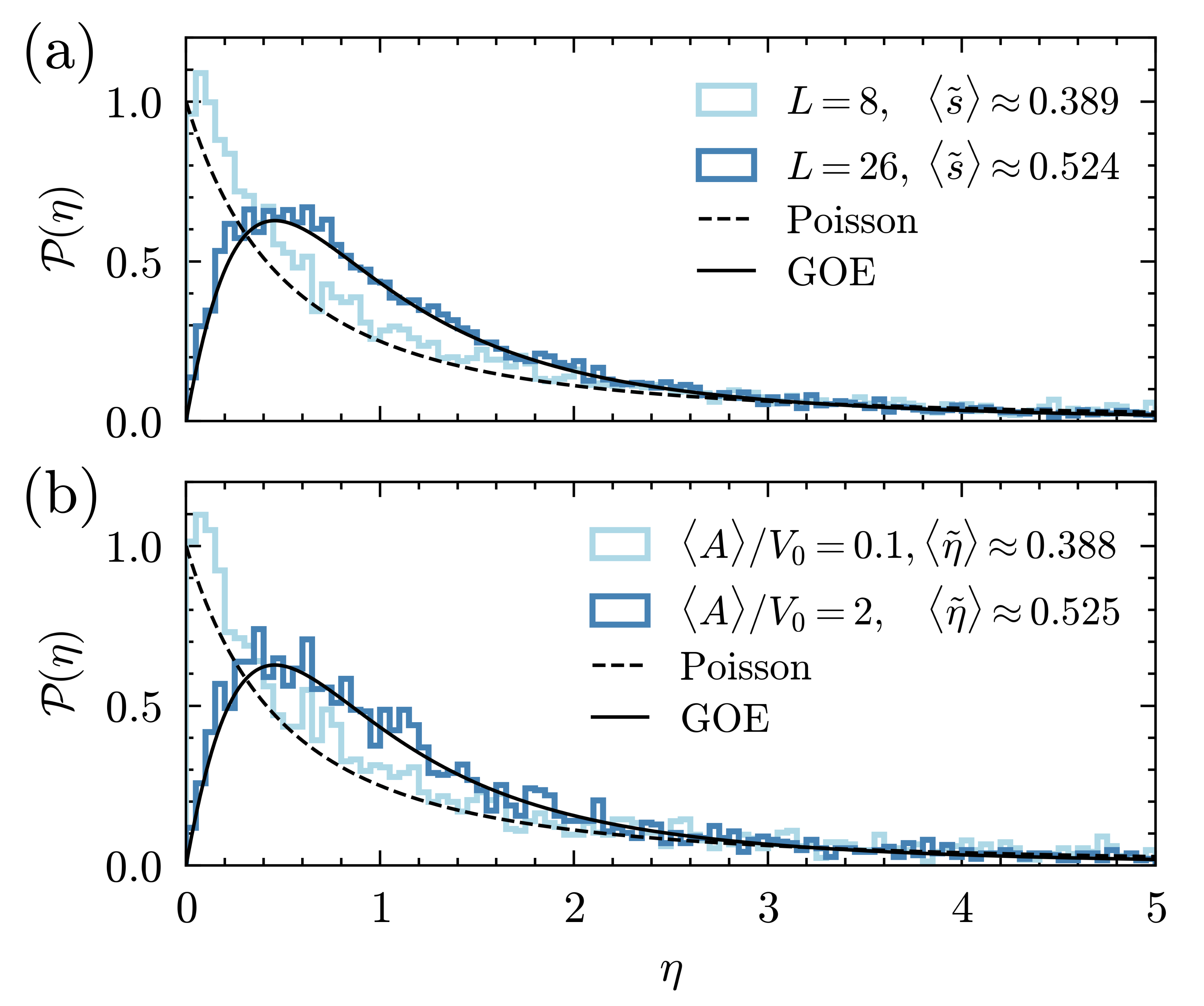}
    \caption{Level-spacing statistics.
    (a) Distribution of consecutive level-spacing ratios for disordered systems of sizes \(L=8\) and \(L=26\). The corresponding symmetrized gap-ratio averages are \(\langle \tilde{\eta} \rangle \approx 0.389\) and \(0.524\), respectively.
    (b) Comparison of weak and strong disorder strengths, \(\langle A \rangle / V_0 = 0.1\) and \(2\), for \(L=10\), yielding \(\langle \tilde{\eta} \rangle \approx 0.388\) and \(0.525\), respectively.
    For reference, the theoretical values are \(\langle \tilde{s} \rangle_{\mathrm{Poisson}} \approx 0.386\) and \(\langle \tilde{\eta} \rangle_{\mathrm{GOE}} \approx 0.536\).}
    
    \label{fig:levelspacing}
\end{figure}

\begin{figure}[t]
    \centering
    \includegraphics[width=\columnwidth]{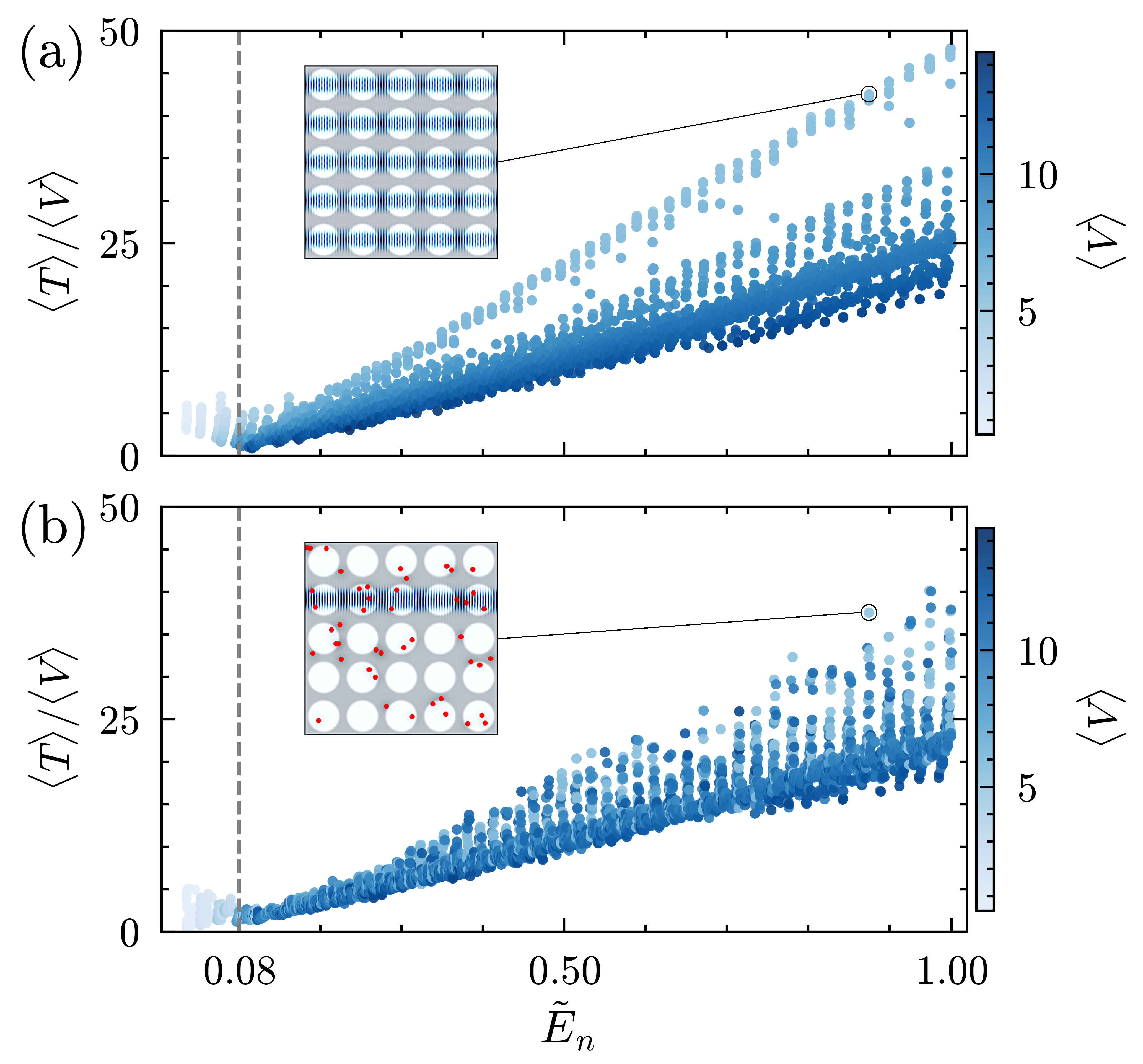}
    \caption{Quantum vs Classical. (a) Ratio of the kinetic and potential energy expectation values \(\langle T \rangle / \langle V \rangle\) as a function of normalized energy in the clean system. High-energy eigenstates show anomalously large ratios, indicating the kinetic-dominated precursors of scarred modes. (b) Upon introducing disorder, degeneracy lifting generates quasi–one-dimensional scarred states that continue to exhibit elevated \(\langle T \rangle / \langle V \rangle\) compared to delocalized, fractal, or ergodic states. Insets illustrate real-space probability densities of the selected eigenstates.}
    \label{fig:cl_vs_QM}
\end{figure}

From the viewpoint of spectral correlations, we quantify the degree of chaos by considering the ratio of consecutive level spacings \(\eta_n=\delta_n/\delta_{n-1}\), where \(\delta_n=E_{n+1}-E_n\). This unfolding-free statistics~\cite{Atas2013} cleanly distinguishes Poisson statistics affiliated with localized spectra from the Wigner-Dyson statistics of the Gaussian orthogonal ensemble (GOE) expected for time-reversal symmetric chaotic systems. We particularly focus on the symmetrized ratio \(\tilde{\eta}_n=\min(\eta_n,1/\eta_n)\), whose mean value provides a compact diagnostic of spectral correlations. In the Poisson and GOE limits, the expected averages are \(\langle \tilde{\eta} \rangle_{\mathrm{Poisson}}\approx0.386\) and \(\langle \tilde{\eta} \rangle_{\mathrm{GOE}}\approx0.536\), respectively. Additional details of our statistics methods are provided in the SM~\cite{SupplementalMaterial}.

Figures~\ref{fig:levelspacing}(a) and (b) summarize the evolution of the level-spacing statistics under varying disorder. In general, one would surmise that strong disorder as well as increasing system size would drive the level statistics toward the Poisson distribution. Surprisingly, our results reveal the opposite trend, in agreement with the behavior observed in the IPR analysis. As shown in Fig.~\ref{fig:levelspacing}(a), the distribution is close to Poissonian for small systems (\(L=8\)). Such behavior agrees with the interpret that weak level correlations arise from localized eigenstates and spectral clustering, thus suppressing level repulsion. Similar effects linked to perturbation-induced variational scarring have previously been reported in 2D quantum wells~\cite{KeskiRahkonen2019b}.

As the system size increases (\(L=26\)), the level statistics unveil a shift toward the Wigner-Dyson form. We see this change as a consequence of the fact that the contribution of scarred states within the relevant spectral window is reduced, even though the total number of AL states continues to grow. Increasing the disorder strength [Fig.~\ref{fig:levelspacing}(b)] produces a similar crossover, where the disappearance of scar-related signatures become apparent in the level statistics. Therefore, for a relevant choice of spectral window and number of levels, we concur that level-spacing statistics can provide indications of the presence of scarred states.

Our results here differ from standard tight-binding predictions, and thus warrant a clarification. In the deep-well limit, where each well hosts a single impurity and Gaussian potentials do not overlap, the disorder reduces to uncorrelated on-site randomness and the continuum model approaches the Anderson tight-binding description (see SM~\cite{SupplementalMaterial} for pedagogical derivation). Only in this regime, a direct comparison is meaningful. In finite tight-binding systems, the localization length \(\xi\) may exceed the system size \(L\), yielding apparently delocalized states that localize asymptotically. By contrast, a smooth confinement and correlated disorder in the continuum support interference effects and scar-like high-energy morphologies that are strongly suppressed in coarse-grained lattice descriptions.

To further elucidate the origin of the continuum-only features, we compare the quantum eigenstates in respect to their classical skeleton. At energies above the confinement barrier, classical trajectories produce smoothly varying fluctuations in the ratio of kinetic \(T\) to potential \(V\) energy. However, Fig.~\ref{fig:cl_vs_QM}(a) disclose that the quantum eigenstates of the clean system deviate from this classical expectation even at energies far above the well depth. The states demonstrate pronounced fluctuations in \(\langle T \rangle / \langle V \rangle\), forming extended stripe-like patterns echoing the underlying potential periodicity. These states dominated by kinetic energy act as an antecedent of scars.

In the presence of disorder [Fig.~\ref{fig:cl_vs_QM}(b)], the degenerate clusters of high-energy sinusoidal states are lifted, and their coherent interference collapses the stripe progenitors into sharply localized quasi–one-dimensional channels, as shown in the inset of Fig.~\ref{fig:cl_vs_QM}(b) [see also Figs.~\ref{fig:crossover}(c-iii \& c-iv)]. These kinetic-energy-dominated outliers correspond to scarred modes arising from interference within groups of degenerate states. The resulting channels are selected variationally, as they extremize the overlap of the eigenstates with the effective impurity landscape, which in a periodic potential is dominated by impurities inside the wells where the wavefunction has appreciable weight. Their persistence under disorder underscores their quantum-interference origin. Rather than fully delocalizing, the high-energy states retain a memory of the potential periodicity and form anisotropic structures absent in coarse-grained Anderson models. This mechanism explains how the interplay of smooth correlated disorder and periodic confinement produces non-ergodic morphologies at energies where classical intuition would anticipate featureless, ergodic dynamics.

\subsection{IV.\quad Summary and Discussion}
We have demonstrated the coexistence of Anderson-localized, extended, and variationally scarred states in finite 2D disordered continua whose bounded geometry supports families of classical POs, arising from finite-size crossovers governed by the rapid growth of the localization length with energy. Although scaling theory rules out a true metallic phase in the large system size limit for short range hoppings, the combination of short but smooth correlated disorder and an underlying periodic confinement produces quasi-localized, anisotropic high-energy modes that have no analogue in typical coarse-grained tight-binding models due to their truncated resolution of the continuum dynamics. 

Rather than a sharp transition, the apparent splitting of spectral ergodicity reflects the interplay between the localization length and the system size, allowing all three types of states to coexist within the same spectral window. The high energy stripe like structures present in the clean system collapse into quasi–one-dimensional channels once local disorder is introduced. Interference among disorder-induced degenerate states leads to directional variational scars unique to finite, periodically structured continua. Collectively, these results reveal the simultaneous presence of conventional Anderson localization and weak ergodicity breaking in realistic mesoscopic systems, with experimentally accessible signatures in electronic, photonic, and cold-atom platforms.

\subsection{V.\quad Acknowledgments}
We acknowledge CSC — Finnish IT Center for Science for computational resources, and Research Council of Finland, ManyBody2D Project No. 349956, for financial support. This project is also supported by the National Science Foundation (Grant No.~2403491).

%%%%%%%%%%%%%%%%%%%%%%%%%%%%%%%%%%%%%%%%%%%%%%

\bibliography{Bibliography}

%%%%%%%%%%%%%%%%%%%%%%%%%%%%%%%%%%%%%%%%%%%%%%

\onecolumngrid
\pagestyle{empty}

\renewcommand{\thefigure}{S\arabic{figure}}
\setcounter{figure}{0}
\renewcommand{\theequation}{S-\arabic{equation}}
\setcounter{equation}{0}

\clearpage

\begin{center}
{\large \bf Coexistence of Anderson Localization and Quantum Scarring in Two Dimensions}

\ \\

{Fartash Chalangari, Anant Vijay Varma, Joonas Keski-Rahkonen and Esa Räsänen}

\ \\

{\large (Supplementary Material)}

\end{center}
%%%%%%%%%%%%%%%%%%%%%%%%%%%%%%%%%%%%%%%%%%%%

In this Supplementary we provide examples of different ergodic and non-ergodic states and their characteristic measures used. Details of the Anderson state tail analysis, comparison of clean and disorder periodic systems with different sizes in a typical scenario. Moreover, for pedagogy a tight-binding derivation assuming one narrow Gaussian bump in each potential well.  

\begin{figure}[H]
    \centering
    \includegraphics[width=0.8\textwidth]{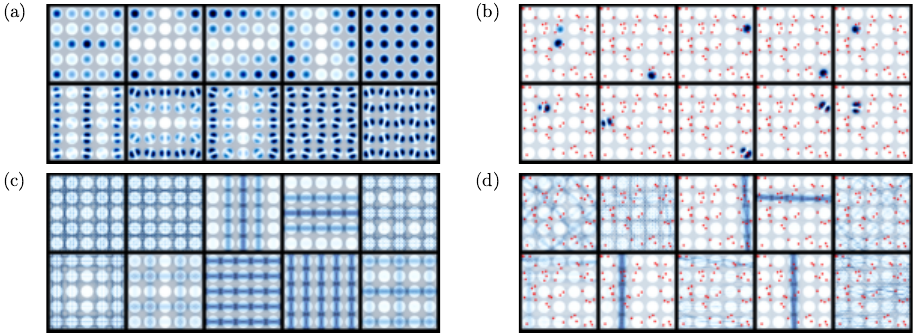}
    \caption{ Typical spectrum hosting multiple species of eigenstates. Selected eigenstates of a system with \(L=10\). White circles mark periodic wells and surrounding gray regions indicate confinement barriers. \textbf{(a)} and \textbf{(c)} are respectively the typical low- and high- energy eigenstates when $ V_{\mathrm{imp}}(x,y)=0$ \textbf{(b)} \& \textbf{(d)} are eigenstates with same eigen indices after inclusion of Gaussian bumps ($ V_{\mathrm{imp}}(x,y)$ is non zero).}
    \label{fig:stateSamples}
\end{figure}

%%%%%%%%%%%%%%%%%%%%%%%%%%%%
\sectA{Scar Metric}

To quantify the degree to which an eigenstate concentrates along lattice rows or columns, we introduce a scar metric based on the probability density
\begin{equation}
\rho(x,y)=|\psi(x,y)|^2,
\end{equation}
with normalization \(\int \rho(x,y)\,dx\,dy=1\).

The periodic geometry defines a set of horizontal and vertical stripe regions aligned with the lattice rows and columns. For each horizontal stripe \(j\) and vertical stripe \(k\), we compute the probability weights
\begin{equation}
\mathcal{P}^{\mathrm{row}}_j=\int_{\mathrm{stripe}\,j}\rho(x,y)\,dx\,dy,
\qquad
\mathcal{P}^{\mathrm{col}}_k=\int_{\mathrm{stripe}\,k}\rho(x,y)\,dx\,dy.
\end{equation}

To measure how strongly a state selects a given channel, we define a local contrast relative to neighboring stripes. For horizontal stripes,
\begin{equation}
\mathcal{C}^{\mathrm{row}}_j=
\mathcal{P}^{\mathrm{row}}_j
-\frac{1}{2}\left(
\mathcal{P}^{\mathrm{row}}_{j-1}
+\mathcal{P}^{\mathrm{row}}_{j+1}
\right),
\end{equation}
and analogously for vertical stripes,
\begin{equation}
\mathcal{C}^{\mathrm{col}}_k=
\mathcal{P}^{\mathrm{col}}_k
-\frac{1}{2}\left(
\mathcal{P}^{\mathrm{col}}_{k-1}
+\mathcal{P}^{\mathrm{col}}_{k+1}
\right).
\end{equation}
Negative values are set to zero, so that only enhanced channel selection contributes to the metric.

The spatial anisotropy of the density within a stripe is characterized by the probability-weighted variances
\begin{equation}
\sigma_x^2=
\frac{\int_{\mathrm{stripe}} (x-\langle x\rangle)^2 \rho(x,y)\,dx\,dy}
{\int_{\mathrm{stripe}} \rho(x,y)\,dx\,dy},
\qquad
\sigma_y^2=
\frac{\int_{\mathrm{stripe}} (y-\langle y\rangle)^2 \rho(x,y)\,dx\,dy}
{\int_{\mathrm{stripe}} \rho(x,y)\,dx\,dy},
\end{equation}
where \(\langle x\rangle\) and \(\langle y\rangle\) are the probability-weighted coordinates evaluated within the same stripe region.

From these variances we define elongation factors
\begin{equation}
\mathcal{A}^{\mathrm{row}}_j=\frac{\sigma_x}{\sigma_y},
\qquad
\mathcal{A}^{\mathrm{col}}_k=\frac{\sigma_y}{\sigma_x},
\end{equation}
so that horizontally elongated states are enhanced in row stripes and vertically elongated states are enhanced in column stripes.

The stripe scores are then defined as
\begin{equation}
S^{\mathrm{row}}_j=\mathcal{C}^{\mathrm{row}}_j\mathcal{A}^{\mathrm{row}}_j,
\qquad
S^{\mathrm{col}}_k=\mathcal{C}^{\mathrm{col}}_k\mathcal{A}^{\mathrm{col}}_k.
\end{equation}
Finally, the scar metric for eigenstate \(n\) is taken as the maximum stripe score over all rows and columns,
\begin{equation}
S_n=\max\!\left(
\max_j S^{\mathrm{row}}_j,\,
\max_k S^{\mathrm{col}}_k
\right).
\end{equation}

Large values of \(S_n\) therefore indicate that the probability density both strongly selects a single lattice channel relative to its neighbors and exhibits pronounced elongation along that direction. In this way, the metric distinguishes channel-like scarred states from compact Anderson-localized states and approximately isotropic extended states. \\

%%%%%%%%%%%%%%%%%%%%%%%%%%%%
\sectA{Tail Analysis}

For an Anderson localized state, the density can be written as Eq.~\ref{eq:density}, and to extract the localization length, we fit \(\ln |\psi(r)|^2\) versus r only in the linear tail window where \(d[\ln A(r)]/dr \approx 0\). In this window, the prefactor is effectively constant, \(A(r)=C\), and from normalization condition of eigenstates, we can derive the relation between \(\mathrm{P}_2\) and localization length:
\begin{equation}
    \begin{split}
    & 1=\int |\psi(r)|^2 \; d^2r = \int_0^\infty 2\pi C \; r \; e^{-r/ \xi} \; dr \\ 
    & \Rightarrow 2\pi C \int_0^\infty re^{-r/\xi} \; dr = 2\pi C \; \xi^2 = 1 \\
    & \Rightarrow C= \frac{1}{2\pi \xi^2}
    \end{split}
    \label{eq:density}
\end{equation}

\begin{equation}
\begin{split}
    & \mathrm{P}_2 = \int |\psi(r)|^4 \; d^2r = \int_0^\infty 2\pi r \; C^2 e^{-2r/\xi} \; dr \\
    & \Rightarrow \mathrm{P}_2 = 2\pi \left(\frac{1}{4 \pi^2 \xi^4}\right) \left(\frac{\xi^2}{4}\right)=\frac{1}{8 \pi} \xi^{-2}
\end{split}
\end{equation}

\begin{figure}[H]
    \centering
    \includegraphics[width=0.5\linewidth]{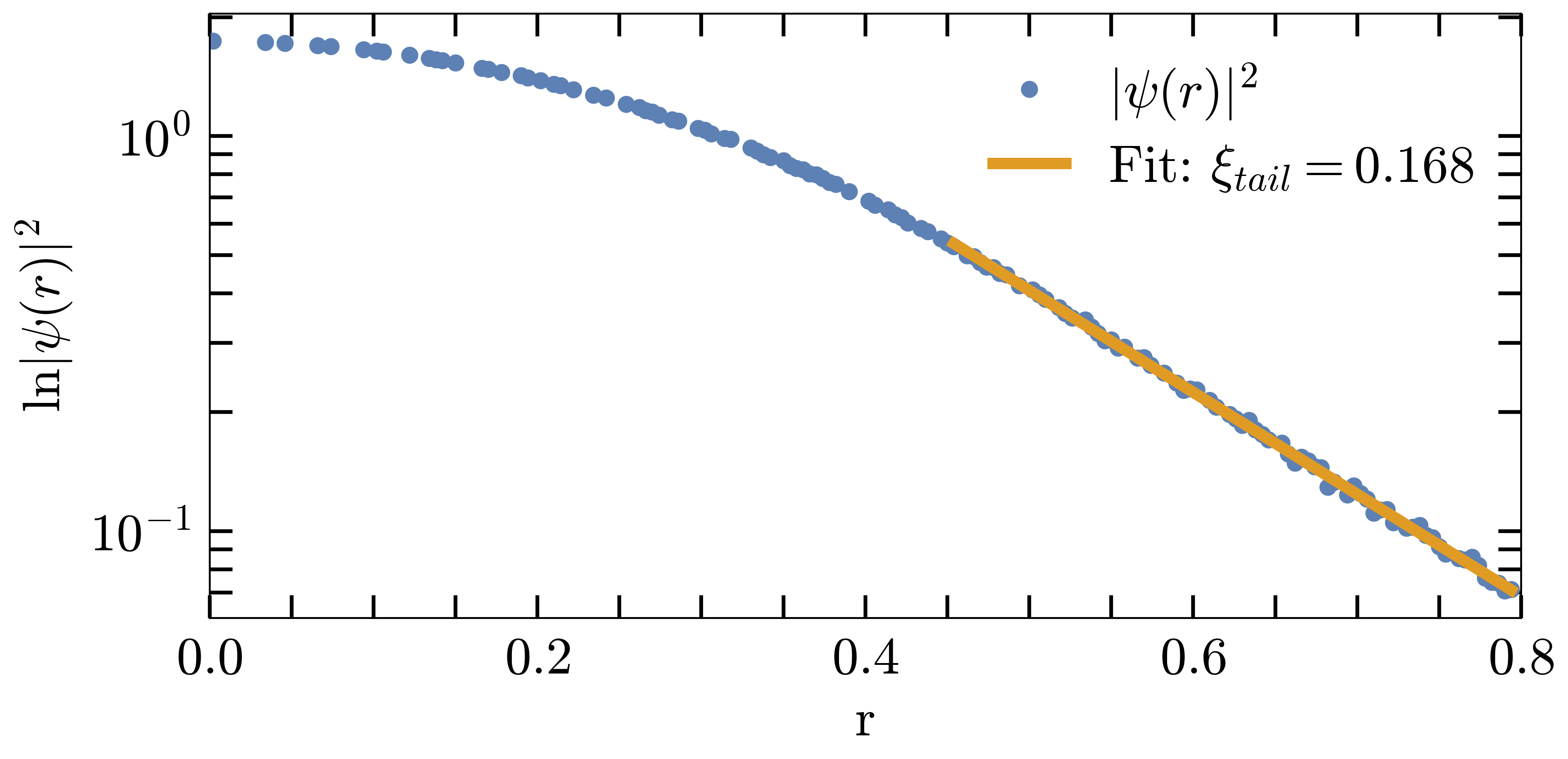}
    \caption{Logarithm of the radial probability density 
    \(\ln |\psi(r)|^2\) of a representative Anderson-localized state 
    as a function of the radial distance \(r\) from the localization center, for a system with \(L=10\) and \(\langle A \rangle / V_0 = 0.3\). The linear behavior on the logarithmic scale confirms the exponential 
    spatial decay characteristic of Anderson localization.}
    \label{fig:exp_decay}
\end{figure}

\begin{figure}[H]
    \centering
    
    % --- Left subfigure ---
    %\begin{minipage}[b]{0.48\linewidth}
     %   \centering
      %  \includegraphics[width=\linewidth]{Figs/ipr_vs_energy.png}
    %\end{minipage}
    %\hfill
    % --- Right subfigure ---
    \begin{minipage}[b]{0.5\linewidth}
        \centering
        \includegraphics[width=\linewidth]{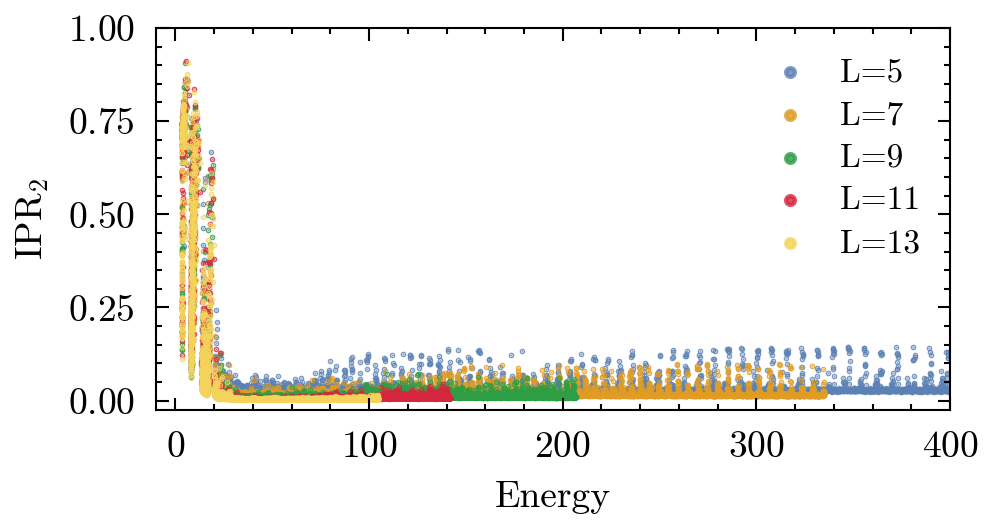}
    \end{minipage}

    % Single caption for the entire figure
    \caption{ 
    %(Left) \(\text{IPR}_2\) versus energy comparing clean and disordered systems, and 
    \(\mathrm{P}_2\) versus energy for different system sizes, showing progressively narrower absolute energy window as the system size increases .}
    \label{fig:ipr-energy-anderson}
\end{figure}

%%%%%%%%%%%%%%%%%%%%%%%%%%%%%%%%%%%%%%%%%%%%%%%%%%%%%

\sectA{Level-spacing ratio statistics}
To characterize spectral correlations we employ the ratio of consecutive level spacings~\cite{Atas2013}. 
Given ordered eigenenergies \(E_n\), the level spacings are defined as
\begin{equation}
\delta_n = E_{n+1}-E_n .
\end{equation}
The ratio of consecutive spacings is then
\begin{equation}
\eta_n = \frac{\delta_n}{\delta_{n-1}} .
\end{equation}

A key advantage of this statistic is that it does not require spectral unfolding. 
This makes it particularly convenient for numerical studies where the density of states 
varies with energy.

In the limit of uncorrelated spectra, characteristic of Anderson-localized systems, 
the ratio distribution follows the Poisson form
\begin{equation}
\mathcal{P}_{\mathrm{Poisson}}(\eta)=\frac{1}{(1+\eta)^2}.
\end{equation}

For time-reversal symmetric chaotic systems the statistics are described by the 
Gaussian orthogonal ensemble (GOE). The corresponding ratio distribution is given by
\begin{equation}
\mathcal{P}_{\mathrm{GOE}}(\eta)=
\frac{27}{8}\frac{\eta+\eta^2}{(1+\eta+\eta^2)^{5/2}} .
\end{equation}

The asymptotic behavior differs for the two cases. For Poisson statistics,
\begin{equation}
\mathcal{P}_{\mathrm{Poisson}}(\eta) \propto \eta^{-2}, \qquad \eta \gg 1,
\end{equation}
while for GOE one finds a faster decay,
\begin{equation}
\mathcal{P}_{\mathrm{GOE}}(\eta) \propto \eta^{-3}, \qquad \eta \gg 1.
\end{equation}
The most pronounced differences, however, occur at small \(\eta\), where GOE statistics exhibit level repulsion, suppressing small spacing ratios, in contrast to the Poisson case. 

For numerical stability it is often convenient to use the symmetrized ratio
\begin{equation}
\tilde{\eta}_n = \min(\eta_n,1/\eta_n),
\end{equation}
which lies in the interval \((0,1)\). The mean value of this quantity provides 
a compact diagnostic of spectral correlations. In the Poisson and GOE limits 
the expected values are
\begin{equation}
\langle \tilde{\eta} \rangle_{\mathrm{Poisson}} \approx 0.386,
\qquad
\langle \tilde{\eta} \rangle_{\mathrm{GOE}} \approx 0.536 .
\end{equation}
These reference values provide a convenient benchmark for identifying the 
degree of spectral correlation in the numerical spectra analyzed in this work.

\begin{figure}
\includegraphics[width=0.5\linewidth]{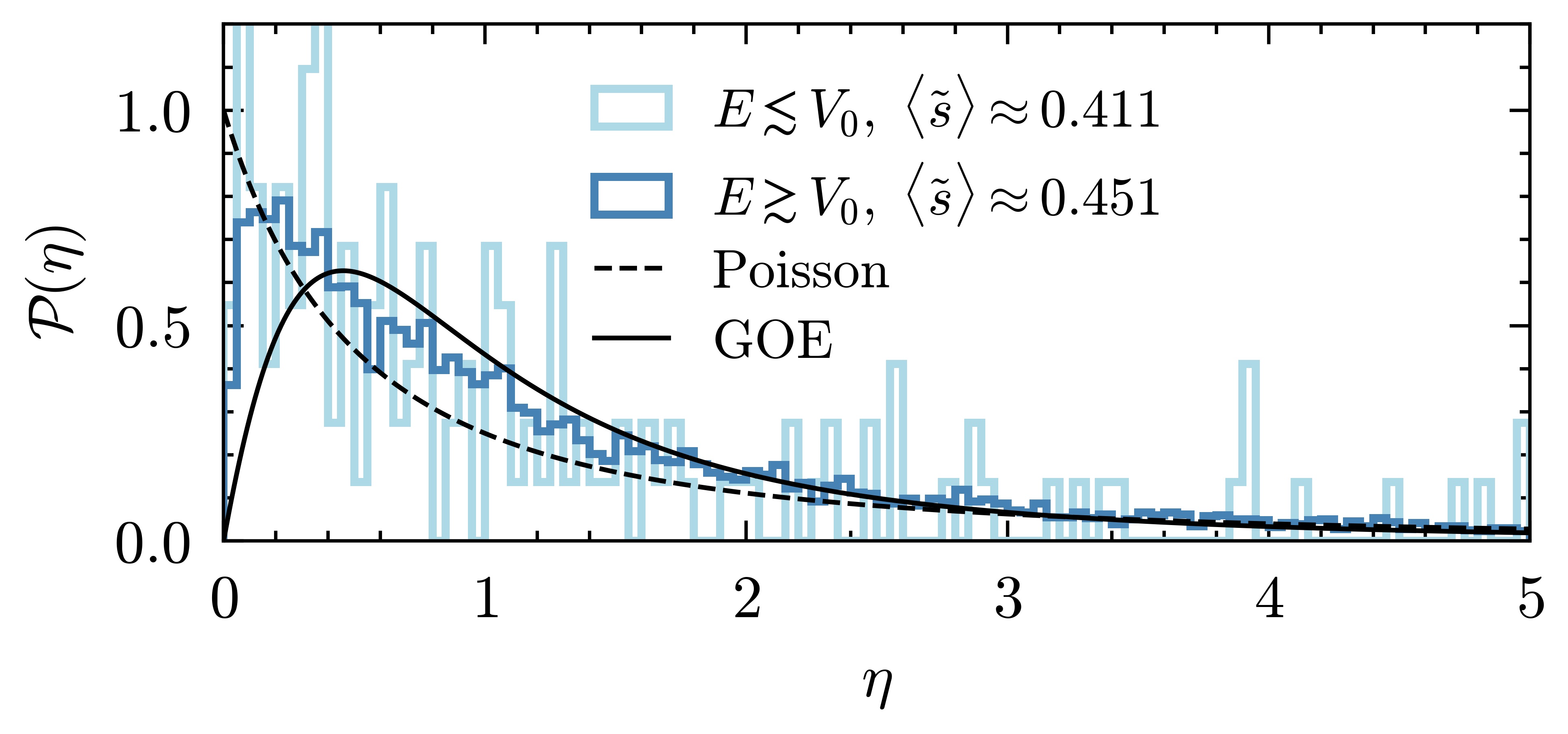}
\caption{
Distribution of the ratio of consecutive level spacings \(\eta\) for eigenstates below and above the barrier height \(V_0\). 
The dashed and solid curves show the Poisson and GOE predictions, respectively. 
Due to the finite number of computed eigenstates, restricting the spectrum to \(E\lesssim V_0\) yields limited statistics, resulting in a noisy histogram that does not fully converge to the Poisson distribution expected for strongly localized states. 
The high-energy sector \(E\gtrsim V_0\) contains a mixture of ergodic extended and scarred eigenstates, producing intermediate spectral statistics between the Poisson and GOE limits.
}
\label{fig:spacing_diffen}
\end{figure}

To further examine spectral correlations, we separate the spectrum into two energy sectors relative to the barrier height \(V_0\). Figure~\ref{fig:spacing_diffen} shows the distribution of the spacing ratio \(\eta\) for eigenstates with energies \(E\lesssim V_0\) and \(E\gtrsim V_0\). 

Because the total number of numerically computed eigenstates is finite, restricting the analysis to low energies significantly reduces the number of available spacing ratios, resulting in limited statistical resolution and a correspondingly noisy histogram. Consequently, the low-energy distribution does not clearly converge to the ideal Poisson form expected for strongly localized spectra.

The high-energy sector contains a mixture of ergodic extended states and anisotropic scarred states. As a result, its spacing statistics lie between the Poisson and GOE limits rather than following a single universal distribution. This mixed character is reflected in the intermediate value of the mean symmetrized spacing ratio \(\langle \tilde{\eta} \rangle\). \\

%%%%%%%%%%%%%%%%%%%%%%%%%%%%%%%%%%%%%%%%%%%%%%%%%%%%%
\sectA{Deriving the Tight-Binding Anderson Model from a Continuum Potential}

We start from the continuum single-particle Hamiltonian confined in a finite region:
\begin{equation}
H = -\frac{\hbar^2}{2m}\nabla^2 + V_{\mathrm{ext}}(\mathbf{r}) + V_{\mathrm{imp}}(\mathbf{r}),
\label{eq:cont_H}
\end{equation}
where $V_{\mathrm{ext}}(\mathbf{r})$ is a periodic array of deep wells and $V_{\mathrm{imp}}(\mathbf{r})$ represents disorder modeled by Gaussian bumps:
\begin{equation}
V_{\mathrm{imp}}(\mathbf{r}) = \sum_{i} A_i \exp\!\left[-\frac{|\mathbf{r}-\mathbf{r}_i|^2}{2\sigma^2}\right].
\label{eq:gaussian_disorder}
\end{equation}

Assume each well in $V_{\mathrm{ext}}$ is deep and narrow, so the lowest eigenstate in each well is strongly localized. Let $\{\phi_i(\mathbf{r})\}$ denote these localized orbitals centered at lattice sites $\mathbf{R}_i$. In the deep-well limit:
\[
\phi_i(\mathbf{r}) \approx \varphi(\mathbf{r}-\mathbf{R}_i), \quad \text{with negligible overlap between distinct wells.}
\]

The full wavefunction is expanded as:
\begin{equation}
\Psi(\mathbf{r}) = \sum_i c_i \phi_i(\mathbf{r}).
\label{eq:expansion}
\end{equation}

The matrix elements in this basis are:
\begin{align}
\langle \phi_i | H | \phi_j \rangle &= \int d^2r\, \phi_i^*(\mathbf{r}) \left[-\frac{\hbar^2}{2m}\nabla^2 + V_{\mathrm{ext}}(\mathbf{r}) + V_{\mathrm{imp}}(\mathbf{r})\right]\phi_j(\mathbf{r}).
\end{align}

For $i=j$ (on-site term):
\begin{equation}
\epsilon_i = \langle \phi_i | H | \phi_i \rangle \approx E_0 + \int d^2r\, |\phi_i(\mathbf{r})|^2 V_{\mathrm{imp}}(\mathbf{r}),
\label{eq:onsite}
\end{equation}
where $E_0$ is the ground-state energy of an isolated well. Since $V_{\mathrm{imp}}$ consists of Gaussian bumps and we assume one bump per well with negligible overlap, the integral picks up only the local impurity amplitude:
\[
\epsilon_i \approx E_0 + A_i.
\]

For $i\neq j$ (hopping term):
\begin{equation}
t_{ij} = \langle \phi_i | H | \phi_j \rangle \approx -\int d^2r\, \phi_i^*(\mathbf{r}) \frac{\hbar^2}{2m}\nabla^2 \phi_j(\mathbf{r}),
\label{eq:hopping}
\end{equation}
which is exponentially small due to negligible overlap. We retain only nearest-neighbor hoppings $t$.

Collecting terms, the effective Hamiltonian in the localized basis is:
\begin{equation}
H_{\mathrm{TB}} = \sum_i \epsilon_i c_i^\dagger c_i + \sum_{\langle i,j\rangle} t \left(c_i^\dagger c_j + c_j^\dagger c_i\right),
\label{eq:TB_H}
\end{equation}
with
\[
\epsilon_i = E_0 + A_i, \quad t \approx \text{constant (nearest-neighbor)}.
\]

Since $A_i$ are random and uncorrelated in the deep-well limit, $\epsilon_i$ are random on-site energies. Thus, Eq.~\eqref{eq:TB_H} is the standard 2D Anderson Hamiltonian:
\begin{equation}
H_{\mathrm{Anderson}} = \sum_i \epsilon_i c_i^\dagger c_i + t \sum_{\langle i,j\rangle} \left(c_i^\dagger c_j + \text{h.c.}\right),
\label{eq:Anderson}
\end{equation}
where $\epsilon_i$ are drawn from a disorder distribution set by the Gaussian amplitudes $A_i$. \\

{\bf \textit{Key Assumptions:}}\\
(1) Deep wells $\Rightarrow$ localized orbitals with negligible overlap. \\
(2) One Gaussian bump per well $\Rightarrow$ uncorrelated on-site disorder.\\ 
(3) Retain only nearest-neighbor hopping $\Rightarrow$ standard TB connectivity. \\ 

This derivation shows that the continuum model with isolated Gaussian impurities reduces to the tight-binding Anderson model in the deep-well limit.

\end{document}